\documentclass[11pt]{article}
\title{Substationarity in Spatial Point Processes}
\author{Tonglin Zhang \footnote{Department of Statistics, Purdue University, 250 North University Street,West Lafayette, IN 47907-2066, Email: tlzhang@purdue.edu} and Jorge Mateu \footnote{Department of Mathematics, Universitat Jaume I, Campus Riu Sec, 12071 Castell\'on, Spain, Email: mateu@mat.uji.es}}
\usepackage{epsfig}
\usepackage{mathrsfs}
\usepackage{amsfonts}
\usepackage{amsmath}
\usepackage{wrapfig,lipsum,booktabs}
\usepackage{bm}
\usepackage{epsfig}
\def\qed{\hfill$\diamondsuit$}
\newtheorem{defn}{Definition}

\newtheorem{thm}{Theorem}
\newtheorem{cor}{Corollary}

\begin{document}
\maketitle
\def\eqalign#1{\null\,\vcenter{\openup\jot\ialign
              {\strut\hfil$\displaystyle{##}$&$\displaystyle{{}##}$
               \hfil\crcr#1\crcr}}\,}

\newpage
\setcounter{page}{1}
\begin{abstract}
The goal of the article is to develop the approach of substationarity to spatial point processes (SPPs). Substationarity is a new concept, which has never been studied in the literature. It means that the distribution of SPPs can only be invariant under location shifts within a linear subspace of the domain. Theoretically, substationarity is a concept between stationariy and nonstationarity, but it belongs to nonstationarity. To formally propose the approach, the article provides the definition of substationarity and an estimation method for the first-order intensity function. As the linear subspace may be unknown, it recommends using a parametric way to estimate the linear subspace and a nonparametric way to estimate the first-order intensity function, indicating that it is a semiparametric approach. The simulation studies show that both the estimators of the linear subspace and the first-order intensity function are reliable. In an application to a forest wildfire data set, the article concludes that substationarity of wildfire occurrences may be assumed along the longitude, indicating that latitude is a more important factor than longitude in forest wildfire studies.

\end{abstract}


\vspace{12pt}

\section{Introduction}
\label{sec:introduction}

The goal of the article is to develop the concept of {\it substationarity} for spatial point processes (SPPs). Substationarity a new concept, which has not been studied in the literature. Theoretically, substationarity can bridge stationarity and nonstationarity, two well-known concepts in the literature of spatial statistics. Substationarity means that the distribution of an SPP is only invariant under any location shift within a linear subspace of the domain. Stationarity means that the distribution is invariant under any location shift within the entire domain. Nonstationarity is the complementary concept of stationarity. It means that the distribution of the SPP can be affected by at least one location shift in the domain. If an SPP is substationary, then its distribution may still be affected by a location shift if it is outside the linear subspace. Therefore, the intersection of substationarity and nonstationarity is not empty. Substationarity provides a way to treat nonstationarity. It can make inferences on nonstationarity easy and convenient. 

The idea of the research is motivated from our recent work on typical events in natural hazards \cite{zhang2014a}. According to its scientific definition, a natural hazard is a naturally occurring event that might have a negative effect on human or environments. Natural hazards include wildfires, tornados, and earthquakes. In our work on forest wildfires, we identified an inhomegenous wildfire pattern in Alberta (Canada) forests. The proportion of large wildfires in the north was higher than that in the south, but the frequency of wildfires in the south was higher than that in the north. Wildfire activities were not significantly affected by their longitude values. It seems that substationarity might be held along the longitude, indicating that it is an important concept in forest wildfire studies.

Statistical approaches to SPPs are important in many scientific disciplines such as forestry \cite{stoyan1994}, epidemiology \cite{benes2005,diggle2006}, wildfires \cite{peng2005,schoenberg2004}, or earthquakes \cite{ogata1988,zhang2014d}. In statistics, an SPP is treated as a pattern of random points developed in an Euclidean space. The number of points within a bounded subset of the Euclidean space is finite. Point distributions and dependence structures are modeled by intensity functions \cite{diggle2003}. The simplifying assumptions of stationarity and isotropy have been developed to make the analysis convenient. Various well-known tools have been proposed. Examples include the $K$-function \cite{ripley1976}, the $L$-function \cite{besag1977}, and the pair correlation function \cite{stoyan1996}. As stationarity is an important assumption, a few methods have been proposed to evaluate it \cite{guan2008,zhang2014c}. Becuase of the concern of the stationarity assumption, recent research often models SPPs under nonstationarity \cite{moller2007,waagepetersen2009}. An important concept called the second-order intensity-reweighted stationarity (SOIRS) has been proposed \cite{baddeley2000}. This concept is powerful in the joint analysis of the first-order and second-order intensity functions under nonstationarity. With the aid of SOIRS, a number of methods for nonstationarity have been proposed \cite{diggle2007,guan2009,guan2010,henrys2009,waagepetersen2007}. SOIRS only specifies the relationship between the first-order and the second-order intensity functions. It does not contain any assumptions related to substationarity, implying that statistical approaches to substationarity can be combined with SOIRs.

The purpose of the article is to develop a formal statistical approach to substationarity in SPPs, including the concept of substationarity and corresponding estimation methods. Since the linear subspace may still be unknown, estimation of the subspace must also be involved. In our approach, we want to estimate the subspace via a parametric way and intensity functions given the linear subspace via a nonparametric way. Therefore, we classify our estimation as a semiparametric approach. The nonparametric component provides the intensity functions given the linear subspace and the parametric component supplies the linear subspace. We evaluate the properties of our estimation methods by simulations and applications. In simulations, we evaluate the performance of the estimators of the linear subspace and the first-order intensity function by studying their mean square error (MSE) values. In applications, we implement our approach to forest wildfire data. We conclude that estimation under substationarity can provide more precise and reliable results than that under nonstationarity. 

To the best of our knowledge, the article is the first one to formally discuss the concept of substationarity. As it has not been previously proposed, it is important to have a formal statistical definition of substationarity at the beginning. Although many research problems can be specified, we only focus on estimation of the first-order intensity functions under substationarity. Many nonparametric or semiparametric methods can be adopted, but we only study the kernel method since it is convenient.

The article is organized as follows. In Section \ref{sec:spatial point processes}, we review the concept of SPPs. In Section \ref{sec:substationarity}, we provide the definition of substationarity, including the evaluation of its theoretical properties. In Section \ref{sec:estimation}, we propose a method to estimate the first-order intensity function under substationarity.  In Section \ref{sec:simulation}, we evaluate the performance of our estimators by Monte Carlo simulations. In Section \ref{sec:application}, we apply our approach to the Alberta forest wildfire data. The paper ends with some discussion in Section \ref{sec:discussion}.

\section{Spatial Point Processes}
\label{sec:spatial point processes}

A spatial point process (SPP) ${\cal N}({\cal S})$ on ${\cal S}$ is composed of random points in a measurable ${\cal S}\subseteq\mathbb{R}^d$. It is treated as the restriction of ${\cal N}$, the SPP on the entire $\mathbb{R}^d$, with points only observed in ${\cal S}$. Therefore, points of ${\cal N}$ in ${\cal S}^c$ (the complementary set of ${\cal S}$) are not observed. Let $\mathscr{B}$ and $\mathscr{B}(A)$ be the collections of Borel sets of $\mathbb{R}^d$ and a measurable $A\subseteq\mathbb{R}^d$, respectively. Let $N(A)$ and $N$ be the numbers of points in $A$ and $\mathbb{R}^d$, respectively. Then, $N(A)$ is finite if $A$ is bounded and $P[N(A)=0]=1$ for any $A\in\mathscr{B}(\mathbb{R}^d)$ with $|A|=0$, where $|A|$ is the Lebesgue measure on $\mathbb{R}^d$.

An SPP ${\cal N}$ is $k$th-order stationary if 
\begin{equation}
\label{eq:kth order stationary}
P[N(A_1)=n_1,\cdots,N(A_l)=n_l]=P[N(A_1+{\bf h})=n_1,\cdots,N(A_l+{\bf h})=n_l]
\end{equation}
for any ${\bf h}\in\mathbb{R}^d$, $l\le k$, $A_1,\cdots,A_l\in\mathscr{B}(\mathbb{R}^d)$, and $n_1,\cdots,n_l\in\mathbb{N}$, where $A+{\bf h}=\{{\bf s}+{\bf h}:{\bf s}\in A\}$. It is strong stationary if (\ref{eq:kth order stationary}) holds for any $l\in\mathbb{N}$. We say ${\cal N}({\cal S})$ is $k$th-order stationary and strong stationary, respectively, if it can be derived by restricting a $k$th-order stationary or a strong stationary ${\cal N}$ on ${\cal S}$.

The $k$th-order intensity function of ${\cal N}$ is defined as 
$$
\lambda_k({\bf s}_1,\cdots, {\bf s}_k)=\lim_{\rho(U_{{\bf s}_i})\rightarrow 0,i=1,\ldots,k}{{\rm E}\{\prod_{i=1}^k N(U_{{\bf s}_i})\}\over \prod_{i=1}^k |U_{{\bf s}_i}|},
$$
where  ${\bf s}_1,\cdots, {\bf s}_k\in\mathbb{R}^d$ are distinct, $U_{\bf s}$ is a neighbor of ${\bf s}$, and $\rho(U_{\bf s})$ is the diameter of $U_{\bf s}$, provided that it almost surely exists in the Lebesgue measure on $\mathbb{R}^d$. If ${\cal N}$ is $k$th-order and strong stationary, respectively, then $\lambda_{l}({\bf s}_1+{\bf h},\cdots,{\bf s}_l+{\bf h})$ is independent of ${\bf h}$ almost surely with respect to the Lebesgue measure on $\mathbb{R}^d$ for any positive $l\le k$ and any $l\in\mathbb{N}$, respectively. 

The mean structure of ${\cal N}$ is 
$$
\mu(A)={\rm E}[N(A)] =\int_A\lambda({\bf s})d{\bf s},$$
where $\lambda({\bf s})=\lambda_1({\bf s})$ is the first-order intensity function. The covariance structure of ${\cal N}$ is 
\begin{equation}
\label{eq:covariance of counts in two events}
\eqalign{
{\rm Cov}[N(A_1),N(A_2)]=&\int_{A_1}\int_{A_2}\{ \lambda_2({\bf s}_1,{\bf s_2})-\lambda({\bf s}_1)\lambda({\bf s}_2)\} d{\bf s_2}d{\bf s_1}+\int_{A_1\cap A_2}\lambda({\bf s})d{\bf s}\cr
=&\int_{A_1}\int_{A_2}\{ g({\bf s}_1,{\bf s_2})-1\} \lambda({\bf s}_1)\lambda({\bf s}_2) d{\bf s_2}d{\bf s_1}+\mu(A_1\cap A_2),
}
\end{equation}
 where $ g({\bf s}_1,{\bf s}_2)=\lambda_2({\bf s}_1,{\bf s}_2)/\{ \lambda({\bf s}_1)\lambda({\bf s}_2)\}$ is the pair correlation function. The covariance function of ${\cal N}$ is 
$$
\Gamma({\bf s}_1,{\bf s}_2)=\{ g({\bf s_1},{\bf s}_2)-1\} \lambda({\bf s}_1)\lambda({\bf s}_2)+\lambda({\bf s}_1)\delta_{{\bf s}_1,{\bf s}_1}({\bf s}_2,{\bf s}_2),$$
 where $\delta_{{\bf s},{\bf s}}$ represents the point measure at $({\bf s},{\bf s})\in\mathbb{R}^d\times\mathbb{R}^d$. By the covariance function, (\ref{eq:covariance of counts in two events}) becomes
\begin{equation}
\label{eq:covariance function}
{\rm Cov}[N(A_1),N(A_2)]=\int_{A_1}\int_{A_2}\Gamma({\bf s}_1,{\bf s}_2)d{\bf s}_2d{\bf s}_1.
\end{equation}
If $g({\bf s}_1,{\bf s}_2)$ only depends on ${\bf s}_1-{\bf s}_2$ or $\|{\bf s}_1-{\bf s}_2\|$ such that it can be expressed as $g({\bf s}_1-{\bf s}_2)$ or $g(\|{\bf s}_1-{\bf s}_2\|)$, then ${\cal N}$ is called a second-order intensity-reweighted stationary (SOIRS) or a second-order intensity-reweighted isotropic (SOIRI) SPP. SOIRS and SOIRI are important concepts for nonstationary SPPs as it can model the first-order and second-order intensity functions together \cite{baddeley2000}. 

If ${\cal N}$ is first-order stationary, then $\lambda({\bf s})=c$ and $\mu(A)=c|A|$ for some $c>0$. If ${\cal N}$ is second-order stationary, then $\lambda({\bf s})=c$, $\mu(A)=c|A|$, $g({\bf s}_1,{\bf s}_2)=g({\bf s}_1-{\bf s}_2)$, 
$${\rm Cov}[{\cal N}(A_1),{\cal N}(A_2)]=c^2\int_{A_1}\int_{A_2}\{g({\bf s}_1-{\bf s}_2)-1\}d{\bf s}_2d{\bf s}_1+c|A_1\cap A_2|$$
and
$${\rm V}[N(A)]=c^2\int_{A}\int_{A} \{g({\bf s}_1-{\bf s}_2)-1\}d{\bf s}_2d{\bf s}_1+c|A|.$$
If ${\cal N}$ is Poisson, then $g({\bf s}_1,{\bf s}_2)=1$, indicating that $V\{N(A)\}={\rm E}[N(A)]$ for any bounded $A\in\mathscr{B}(\mathbb{R}^d)$. Only the mean structure is important in Poisson SPPs. However, both the mean and variance structures are important in non-Poisson SPPs. 


\section{Substationarity}
\label{sec:substationarity}

The main purpose of this section is to provide the formal definition of substationarity as well as corresponding properties. As substationarity is a new concept which has not been studied in the literature before, it is also important to provide asymptotic theory under substationarity. The theory are useful in the evaluation of theoretical properties of estimators provided in the next section.  

\begin{defn}
\label{defn:definition of substationarity}
We say ${\cal N}$ is $k$th-order substationary in a linear subspace ${\cal L}\subseteq\mathbb{R}^d$ if (\ref{eq:kth order stationary}) holds for any ${\bf h}\in{\cal L}$, $l\le k$, $A_1,\cdots,A_l\in\mathscr{B}(\mathbb{R}^d)$, and $n_1,\cdots,n_l\in\mathbb{N}$. We say ${\cal N}$ is strong substationary in ${\cal L}$ if it is $k$th-order substationary in ${\cal L}$ for any $l\in\mathbb{N}$. For any ${\cal S}\subseteq\mathbb{R}^d$, we say ${\cal N}({\cal S})$ is $k$th-order substationary or strong substationary in ${\cal L}$ or ${\cal L}\cap {\cal S}$ equivalently if ${\cal N}({\cal S})$ can be restricted by a $k$th-order substationary or strong substationary ${\cal N}$ in ${\cal L}$ on ${\cal S}$. 
\end{defn}

Obviously, if ${\cal N}$ is $k$th-order substationary and its $k$th-order intensity function almost surely exists, then 
\begin{equation}
\label{eq:kth substationary}
\lambda_l({\bf s}_1,\cdots,{\bf s}_l)=\lambda_l({\bf s}_1+{\bf h},\cdots,{\bf s}_l+{\bf h})
\end{equation}
 almost surely with respect to the Lebesgue measure of $\mathbb{R}^d$ for any ${\bf h}\in {\cal L}$, $l\le k$, and distinct ${\bf s}_1,\cdots,{\bf s}_l\in\mathbb{R}^d$. If ${\cal N}$ is $k$th-order substationary in ${\cal L}$, then it is also $k$-th order substationary in any linear subspace ${\cal L}'\subseteq{\cal L}$. Therefore, the linear subspace ${\cal L}$ in Definition \ref{defn:definition of substationarity} is generally not unique. 

\begin{defn}
\label{defn:definition of intrinsic substationarity}
We say ${\cal N}$ is $k$th-order intrinsically substationary or intrinsically strong substationary in ${\cal L}$ if it is substationary or strong substationary in ${\cal L}$ but not in any linear subspace ${\cal L}'$ of $\mathbb{R}^d$ satisfying ${\cal L}\subseteq{\cal L}'$ but ${\cal L}\not={\cal L}'$. We say ${\cal N}({\cal S})$ is $k$th-order intrinsically substationary or intrinsically strong substationary in ${\cal L}$ or ${\cal L}\cap{\cal S}$ equivalently if it can be restricted by a $k$th-order intrinsically substationary or intrinsically strong substationary in ${\cal L}$ on ${\cal S}$.
\end{defn}

If ${\cal N}$ is substarionary in both ${\cal L}_1$ and ${\cal L}_2$, then (\ref{eq:kth order stationary}) holds for any ${\bf h}_1\in{\cal L}_1$ and ${\bf h}_2\in{\cal L}_2$. For any ${\bf h}\in{\rm span}\{{\cal L}_1,{\cal L}_2\}$, there exist ${\bf h}_1\in{\cal L}_1$ and ${\bf h}_2\in{\cal L}_2$ such that ${\bf h}={\bf h}_1+{\bf h}_2$. For any $l\le k$, we have
$$\eqalign{
P[N(A_1+{\bf h})=n_1,\cdots,N(A_l+{\bf h})=n_l]=&P[N(A_1+{\bf h}_1+{\bf h}_2)=n_1,\cdots,N(A_l+{\bf h}_1+{\bf h}_2)=n_l]\cr
=&P[N(A_1+{\bf h}_1)=n_1,\cdots,N(A_l+{\bf h}_1)=n_l]\cr
=&P[N(A_1)=n_1,\cdots,N(A_l)=n_l],\cr
}$$
implying that ${\cal N}$ is also substationary in ${\rm Span}\{{\cal L}_1,{\cal L}_2\}$. Thus, the linear subspace ${\cal L}$ in Definition \ref{defn:definition of intrinsic substationarity} is unique. A $k$th-order intrinsically substationary ${\cal N}$ in ${\cal L}$ is $k$th-order stationary if and only if ${\cal L}=\mathbb{R}^d$. If ${\cal N}$ is intrinsically substationary in ${\cal L}$, then it is substationary in any linear subspace ${\cal L'}$ of ${\cal L}$ but not in any linear subspace ${\cal L}'$ of $\mathbb{R}^d$ strictly covering ${\cal L}$.

\begin{figure}
\centerline{\includegraphics[angle=270,width=80mm]{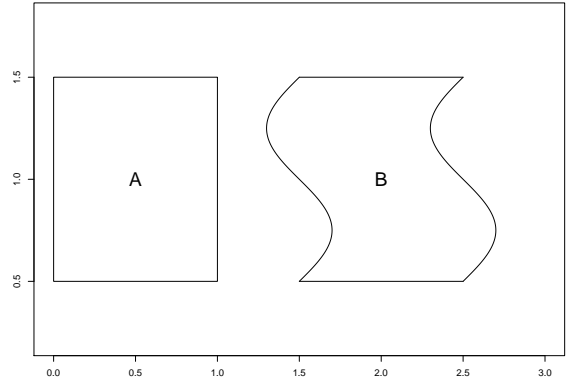}}
\caption{\label{fig:mean structures of two sets under substationarity} Equality of expected counts in two subsets under substationarity along the horizontal axis}
\end{figure}

If ${\cal N}$ is substationary in ${\cal L}$, then for any ${\bf h}\in{\cal L}$ there is $\mu(A)=\mu(A+{\bf h})$. This statement can be true in a more general case. Suppose ${\cal N}$ is substationary in the horizontal axis of $\mathbb{R}^2$ (i.e., $d=2$) such that ${\cal L}=\{(x,0):x\in\mathbb{R}\}$. Then, the first-order intensity of ${\cal L}$ only depends on the vertical value of the point, indicating that we can express $\lambda({\bf s})=\lambda(y)$ for any ${\bf s}=(x,y)\in\mathbb{R}^2$. Let $\nu_r$ be the Lebesgue measure on $\mathbb{R}^r$. For any $A\in\mathbb{R}^2$, there is 
$$\mu(A)=\int_{-\infty}^\infty \lambda(y)\nu_1(A_y)dy,$$
where $A_y=\{{\bf s}=(x,y):(x,y)\in A\}$. For any measurable bounded $A,B\subseteq\mathbb{R}^2$, we may still have $\mu(A)=\mu(B)$ even if $B\not=A+{\bf h}$ for any ${\bf h}\in{\cal L}$ (e.g., the case displayed in Figure \ref{fig:mean structures of two sets under substationarity}).  We summarize this issue into the following theorems.

\begin{thm}
\label{thm:equality of two sets in a partition}
Let ${\cal N}$ be substationary in ${\cal L}\subseteq \mathbb{R}^d$. For any measurable bounded $A,B\in\mathbb{R}^2$, if there exist a partition $\{A_1,A_2,\cdots\}$ of $A$ and a partition $\{B_1,B_2,\cdots\}$ of $B$ such that for every $i$ there exists ${\bf h}_i\in{\cal L}$ satisfying $B_i=A_i+{\bf h}_i$, then $\mu(A)=\mu(B)$.
\end{thm}

\noindent
{\bf Proof:} Straightforwardly, there is 
$$\mu(A)=\sum_{i=1}^\infty \mu(A_i)=\sum_{i=1}^\infty \mu(A_i+{\bf h}_i)=\sum_{i=1}^\infty \mu(B_i)=\mu(B).$$
Then, we draw the conclusion. \qed

\begin{thm}
\label{thm:equality of two sets}
Let ${\cal N}$ be substationary in ${\cal L}\subseteq \mathbb{R}^d$. For any measurable bounded $A,B\in\mathbb{R}^2$, if $\nu_r(A_{\bf v})=\nu_r(B_{\bf v})$ almost surely for any ${\bf v}\in\mathbb{R}^d$, where $A_{\bf v}=\{{\bf s}\in A: {\bf s}-{\bf v}\in {\cal L}\}$ and $r$ is the dimension of ${\cal L}$, then ${\rm E}[N(A)]={\rm E}[N(B)]$.
\end{thm}

\noindent
{\bf Proof:} Let ${\bf u}_1,\cdots,{\bf u}_d$ be the orthogonal bases of $\mathbb{R}^d$, where the previous $r$ vectors form the orthogonal bases of ${\cal L}$.  Let ${\cal L}^\perp=\{{\bf v}\in\mathbb{R}^d:{\bf v}=\sum_{i=r+1}^d x_i{\bf u}_i,x_i\in\mathbb{R}\}$ be the orthogonal space of ${\cal L}$ in $\mathbb{R}^d$ . Let ${\bf s}_{\cal L}$ and ${\bf s}_{{\cal L}^\perp}$ be the orthogonal projection of ${\bf s}$ on ${\cal L}$ and ${\cal L}^\perp$, respectively. Then, the first-order intensity function of ${\cal N}$ can be expressed as $\lambda({\bf s})=\lambda({\bf s}_{{\cal L}^\perp})$ for any ${\bf s}\in A$. We have
$$\eqalign{
\mu(A)=&\int_{{\bf s}\in A}\lambda({\bf s})d{\bf s}\cr
=&\int_{{\cal L}^\perp}\lambda({\bf s}_{{\cal L}^\perp})\nu_r(A_{{\bf s}_{{\cal L}^\perp}})d{\bf s}_{{\cal L}^\perp}\cr
=&\int_{{\cal L}^\perp}\lambda({\bf s}_{{\cal L}^\perp})\nu_r(B_{{\bf s}_{{\cal L}^\perp}})d{\bf s}_{{\cal L}^\perp}\cr
=&\int_{{\bf s}\in B}\lambda({\bf s})d{\bf s}\cr
=&\mu(B).
}$$ 
We draw the conclusion. \qed

\bigskip
Theorems \ref{thm:equality of two sets in a partition} and \ref{thm:equality of two sets} can be used to study the relationship between expected numbers of counts between two regions. It is not enough to use them to study their joint distribution. As it depends on types of ${\cal N}$, we study the properties of the joint distribution under the framework of asymptotics. Let $A_{z,{\cal L}}=\{{\bf v}+z{\bf u}:{\bf v}+{\bf u}\in A,{\bf v}\in{{\cal L}}^{\perp},{\bf u}\in{\cal L}\}$ and $A_{{\bf v},z,{\cal L}}=\{{\bf s}\in A_{z,{\cal L}}:{\bf s}-{\bf v}\in{\cal L}\}$ for any $A\in\mathscr{B}(\mathbb{R}^d)$, where ${\cal L}$ is a linear subspace of $\mathbb{R}^d$.  Then, $A_{{\bf v},z,{\cal L}}=\{{\bf v}+z{\bf u}:{\bf v}+{\bf u}\in A,{\bf v}\in{{\cal L}}^\perp,{\bf u}\in{\cal L}\}$ and $\nu_{r}(A_{{\bf v},z,{\cal L}})=z^r\nu_{r}(A_{{\bf v},1,{\cal L}})$. If ${\cal N}$ is substationary in ${\cal L}$ and $A$ is bounded, then 
$$\eqalign{
\mu(A_{z,{\cal L}})=&\int_{{\bf s}\in A_{z,{\cal L}}} \lambda({\bf s})d{\bf s}\cr
=&\int_{{{\cal L}}^\perp}\lambda({\bf s}_{{\cal L}^\perp})\nu_{r}(A_{{\bf s}_{{{\cal L}}^\perp},z,{\cal L}})d{\bf s}_{{{\cal L}}^\perp}\cr
=&z^{r}\int_{{{\cal L}}^\perp}\lambda({\bf s}_{{\cal L}^\perp})\nu_{r}(A_{{\bf s}_{{{\cal L}}^\perp},1,{\cal L}})d{\bf s}_{{{\cal L}}^\perp}\cr
=&z^{r}\mu(A).
}$$
If ${\cal N}$ is Poisson, then ${\rm V}[N(A_{z,{\cal L}})]=\mu(A_{z,{\cal L}})=z^r\mu(A)$ and 
$$M_{z,{\cal L}}(A)=z^{-{r\over 2}}[N(A_{z,{\cal L}})-\mu(A_{z,{\cal L}})]\stackrel{D}\rightarrow N[0,\mu(A)]$$
 as $z\rightarrow\infty$. 

Let ${\cal A}$ be a collection of Borel sets of $\mathbb{R}^d$. Let ${\cal A}_{z,{\cal L}}$, $N({\cal A}_{z,{\cal L}})$, $\mu({\cal A}_{z,{\cal L}})$ be vectors composed of $A_{z,{\cal L}}$, $N(A_{z,{\cal L}})$, and $\mu(A_{z,{\cal L}})$ for all $A\in{\cal A}$, respectively. If ${\cal A}$ is a finite collection of disjoint subsets such that it can be expressed as ${\cal A}=\{A_1,\cdots,A_m\}$ with disjoint $A_1,\cdots,A_m$, then
\begin{equation}
\label{eq:limiting distribution of many sets}
M_{z,{\cal L}}({\cal A})\stackrel{D}\rightarrow N[0,{\rm diag}(\mu({\cal A}))],
\end{equation}
where $M_{z,{\cal L}}({\cal A})$ is the vector composed of $M_{z,{\cal L}}(A)$ for all $A\in{\cal A}$.

For any $V\in\mathscr{B}({{\cal L}}^\perp)$, let 
\begin{equation}
\label{eq:defintion of collection in functional limit theorm}
A_{{\bf t},V}=(0,t_1{\bf u}_1]\times\cdots\times(0,t_{r}{\bf u}_{r}]\times V, 
\end{equation}
where $t_i>0$, $(0,t_i{\bf u}_i]=\{{\bf s}=x{\bf u}_i:0<x\le t_i\}$, and ${\bf u}_1,\cdots,{\bf u}_{r}$ are the orthogonal bases of ${\cal L}$. Then,
\begin{equation}
\label{eq:limiting distribution of bivariate sets}
 \left(\begin{array}{c} M_{z,{\cal L}}(A_{{\bf t},V}) \cr M_{z,{\cal L}}(A_{{\bf t}',V})\end{array}\right)\stackrel{D}\rightarrow N\left[\left(\begin{array}{c} 0 \cr 0 \end{array}\right), \left(\begin{array}{cc} \mu(A_{{\bf t},V}) & \mu(A_{{\bf t}\wedge{\bf t}',V}) \cr \mu(A_{{\bf t}\wedge{\bf t}',V})  & \mu(A_{{\bf t}',V})\cr\end{array}\right)\right],
\end{equation}
as $z\rightarrow\infty$. The finite-dimensional central limit theorem of $N({\cal A}_{z,{\cal L}})$ can be derived by (\ref{eq:limiting distribution of many sets}) and (\ref{eq:limiting distribution of bivariate sets}), but it is not enough for us to study properties of the estimator of the first-order intensity proposed in the next section. To study the properties, we need the functional central limit theorem of $M_{z,{\cal L}}({\cal A})$ when ${\cal A}$ contains infinitely number of measurable subsets of $\mathbb{R}^d$. A typical way to show functional central limit theorem is to combine the finite-dimensional asymptotics with the tightness \cite{whitt2007}. A typical way to prove the tightness is the evaluation of the bracketing entropy number, which is used in the following theorem.

\begin{thm}
\label{thm:functional central limit theorem for Poisson substationary SPP}
Let ${\cal N}$ be a Poisson substationary SPP in ${\cal L}$. If ${\cal A}_V=\{A_{{\bf t},V}: {\bf t}=(t_1,\cdots,t_{r})\in[0,\infty)^{r}\}$ for some $V\subseteq\mathscr{B}({{\cal L}}^\perp)$, then $M_{z,{\cal L}}({\cal A}_V)$ weakly converges to a mean zero Gaussian random field on $[0,\infty)^{r}$ with the covariance structure given by the right side of (\ref{eq:limiting distribution of bivariate sets}).
\end{thm}

\noindent
{\bf Proof:} We show the conclusion by the standard empirical process approach.  Let ${\cal A}_{V,{\bf a}}=\{A_{{\bf t},V}: {\bf t}=(t_1,\cdots,t_{r})\in[0,a_1]\times\dots\times\prod_{i=1}^{r} [0,a_i]\}$ for any ${\bf a}=(a_1,\cdots,a_{r})^\top\in (0,\infty)^{r}$. Let $F({\bf t})=\mu(A_{{\bf t},V})/\mu(A_{{\bf a},V})$ for any ${\bf t}\preceq {\bf a}$. Then, $F$ is an $r$-dimensional marginal uniformly distributed CDF on the $\sigma$-field generated by ${\cal A}_{{\bf a},V}$. Let $F_i$ be the $i$th CDF of $F$. For any $\epsilon\in(0,1)$, there is an integer $J$ such that ${r}/\epsilon^2\le J\le r/\epsilon^2+1$. Let $x_{ij}=ja_i/(J+1)$ for $j=0,1,\cdots,J+1$. Then, $\epsilon^2/(\epsilon^2+r)\le F_i(x_{i(j+1)})-F_i(x_{ij})\le \epsilon^2/r$. Let $X_{\epsilon}=\{{\bf x}=(x_1,\cdots,x_{r}): x_i=x_{ij}\ {\rm for\ some}\ j=0,1,\cdots,J+1\}$. Then, $\#X_\epsilon=(J+2)^{r}\le [(r+3)/\epsilon^2]^{r}$. For any $g_{\bf x}\in{\cal G}=\{I_{{\bf x}}: {\bf x}\in\prod_{i=1}^{r}[0,a_i]\}$, we can find ${\bf x}',{\bf x}''\in X_\epsilon$ such that ${\bf x}'\preceq {\bf y}\preceq{\bf x}''$ but there is no ${\bf x}^*\in X_\epsilon$ satisfying $x_i'<x_i^*< x_i''$ for some $i=1,\cdots,r$, where $x_i$, $x_i^*$, and $x_i''$ are the $i$th component of ${\bf x}$, ${\bf x}^*$, and ${\bf x}''$, respectively. Then, $g_{{\bf x}'}\le g_{\bf x}\le g_{{\bf x}''}$ and 
$$\|g_{{\bf x}''}-g_{{\bf x}'}\|_{F}^2=\int_{\prod_{i=1}^{r}[0,a_i]}|g_{{\bf x}''}({\bf x})-g_{{\bf x}'}({\bf x})|^2F(d{\bf x})\le \sum_{i=1}^{r}[F_i(x_i'')-F_i(x_i')]\le\epsilon^2.$$
Because 
$$\int_0^1 \log^{1/2}(\#X_\epsilon)d\epsilon\le\int_0^1 \{r[\log(r+3)+2\log\epsilon]\}^{1/2}d\epsilon <\infty,$$ 
we conclude that ${\cal G}$ is $F$-Donsker \cite[P. 270]{vandervaart1998}, implying that the conclusion holds in $\prod_{i=1}^{r}[0,a_i]$ for any ${\bf a}\in(0,\infty)^{r}$. We draw the conclusion of the theorem by letting $a_i\rightarrow\infty$ for all $i$. \qed   

\bigskip
 Theorem \ref{thm:functional central limit theorem for Poisson substationary SPP} supplies the functional central limit theorem of $M_{z,{\cal L}}({\cal A})$ if ${\cal N}$ is Poisson, but it does not provide any similar result of $M_{z,{\cal L}}({\cal A})$ if it is not. A critical issue in the case when ${\cal N}$ is non-Poison is the presence of dependence structures. In particular, for any disjoint $A$ and $B$, if ${\cal N}$ is Poisson, then $N(A)$ and $N(B)$ are independent Poisson random variables with expected values $\mu(A)$ and $\mu(B)$, respectively. If ${\cal N}$ is not Poisson, then the dependence between $N(A)$ and $N(B)$ must be addressed. This requires us to study the property of the second-order intensity function.

Let $A$ and $B$ be bounded measurable subsets of $\mathbb{R}^d$. For any ${\bf h}\in{\cal L}$, there is
$$\eqalign{
{\rm Cov}[N(A+{\bf h}),N(B)]=&\int_{A+{\bf h}}\int_{B}\Gamma({\bf s}_1,{\bf s}_2)d{\bf s}_2d{\bf s}_1\cr
=&\int_{A}\int_{B}[g({\bf s}_1-{\bf s}_2-{\bf h})-1]\lambda({\bf s}_1)\lambda({\bf s}_2)d{\bf s}_2d{\bf s}_1+\int_{(A+{\bf h})\cap B}\lambda({\bf s})d{\bf s}.\cr
} $$
If $\|{\bf h}\|$ is large such that $(A+{\bf h})\cap B=\phi$, then 
$${\rm Cov}[N(A+{\bf h}),N(B)]=\int_{A}\int_{B}[g({\bf s}_1-{\bf s}_2-{\bf h})-1]\lambda({\bf s}_1)\lambda({\bf s}_2)d{\bf s}_2d{\bf s}_1.$$
If $g({\bf s}_1-{\bf s}_2-{\bf h})\rightarrow 1$ as $\|{\bf h}\|\rightarrow\infty$, then ${\rm Cov}[N(A+{\bf h}),N(B)]\rightarrow 0$, indicating that $N(A+{\bf h})$ and $N(B)$ are almost independent. To theoretically address this issue, we need to assume that ${\cal N}$ satisfies the strong mixing condition. This approach was first introduced for dependent random variables by \cite{rosenblatt1956} and later extended to stationary SPPs by \cite{ivanoff1982}. Here we want to modify it to substationarity SPPs. 

Suppose ${\cal N}$ is substationarity in ${\cal L}$. Let $\mathscr{B}(A)$ be the collection of Borel sets generated by $A$. Denote the diameter of $A$ by $\rho(A)$ and $\rho(A_1,A_2)$ as the minimum distance between $A_1$ and $A_2$, where $\rho(A)=\sup_{{\bf s},{\bf s}'\in A}\|{\bf s}-{\bf s}'\|$ and $\rho(A_1,A_2)=\min_{{\bf s}\in A_1,{\bf s}'\in A_2}\|{\bf s}-{\bf s}'\|$. Let 
$$\eqalign{
\alpha(u,v)=\sup\{|P&(U_1\cap U_2)-P(U_1)P(U_2)|:U_1\in\mathscr{B}(A_1),U_2\in\mathscr{B}(A_2),\cr
& \rho(A_1,A_2)\ge u, \rho(A_1)\le v,\rho(A_2)\le v, A_1,A_2\in\mathscr{B}(\mathbb{R}^d)\}
}
$$
be the mixing coefficients, where $P(U)$ is the distribution of $N(U)$. We say ${\cal N}$ is {\it strongly mixing} if $\alpha(zu,zv)\rightarrow 0$ as $z\rightarrow\infty$. 

We want to derive the functional central limit theorem of $M_{z,{\cal L}}({\cal A}_{{\bf t},V})$ for ${\bf t}\in[0,\infty)^{r}$ and $V\in\mathscr{B}({{\cal L}}^\perp)$. Our proof is based on a classical way. It  was initially introduced by \cite{ibragimov1962} and later modified by \cite{herrndorf1984}. The main idea is to split $A_{z,{\cal L}}$ for $A\in{\cal A}_{{\bf t},V}$ into two components $B$ and $C$. Both $B$ and $C$ can be writing into the sum of blocks, where counts in blocks of $B$ are almost independent and counts in blocks of $C$ can be igrnored. This is a popular idea in the proof of the asymptotic normality for stationary time series, which can also be used to SPPs. Since the proof of our functional central limit theorm is just a simple usage of the popular idea, we decide to only briefly display it.

\begin{thm}
\label{thm:functional central limit theorem strong mixing}
Assume ${\cal N}$ is strongly mixing and substationary in ${\cal L}$. If the fourth intensity function of ${\cal N}$ is uniformly bounded and
\begin{equation}
\label{eq:mixing coefficient order}
\int_0^\infty z^{d-{1\over 2}}\alpha(zu,zv)dz<\infty
\end{equation}
for any $u$ and $v$, then $M_{z,{\cal L}}({\cal A}_{V})$ weakly converges to a Gaussian process with independent increments. 
\end{thm}

\noindent
{\bf Proof:} Let $A_i=U_i\times V$ For any disjoint $U_1,\cdots,U_m\in\mathscr{B}({\cal L})$. Define ${\cal A}=\{A_1,\cdots,A_m\}$. Using the method in Theorem 1.3 of \cite{ibragimov1962}, we can partition ${\cal A}$ into many small blocks, denoted by ${\cal B}=\{{\cal B}_{1},\cdots,{\cal B}_{k_1}\}$ and ${\cal C}=\{C_{1},\cdots,{\cal C}_{k_2}\}$, where $k_1,k_2\rightarrow\infty$ as $z\rightarrow\infty$, such that 
$$\min_{B\in{\cal B}_{j,n},B'\in{\cal B}_{j',n},j\not=j'}\rho(B,B')\ge u$$
 and $N({\cal A}_{z,{\cal L}})=N({\cal B}_{z,{\cal L}})+N({\cal C}_{z,{\cal L}})$. By the method of Theorem 1.4 in \cite{ibragimov1962}, we can choose $k_1$ such that it is bounded by $z^{(1+u)/(2d)}$ for any positive $u$ if $z$ is sufficiently large. Then, there is 
$$\left|{\rm E}e^{it\sum_{j=1}^m M_{z,{\cal L}}(A_j)}-\prod_{j=1}^{k_1}{\rm E}e^{it M_{z,{\cal L}}({\cal B}_{j})}\right|\le 4k_1\alpha(zu,zv),$$
where $v=\max(\rho(U_i))$. If (\ref{eq:mixing coefficient order}) holds, then the right side of the above goes to $0$ as $z\rightarrow\infty$. Since $\lambda_4$ is uniformly bounded, we conclude that the Lyapounov Condition \cite[P. 362]{billingsley1995} holds, implying that the asymptotic normality holds. We draw the conclusion about the central limit theorem of $M_{z,{\cal L}}({\cal A})$ for finite ${\cal A}$. By the same method in the proof of the tightness that we have displayed in Theorem \ref{thm:functional central limit theorem for Poisson substationary SPP}, we can show the tightness of the distribution of $M_{z,{\cal L}}({\cal A}_{V})$ for sufficiently large $z$. Then, we draw the functional central limit theorem for $M_{z,{\cal L}}({\cal A}_{V})$, implying the conclusion of the theorem. \qed

\begin{cor}
\label{cor:functional central limit theorem strong mixing}
If all conditions of Theorem \ref{thm:functional central limit theorem strong mixing} hold, then there exists $C>0$ such that for any $A\in\mathscr{B}(\mathbb{R}^d)$ there is $M_{z,L}(A)\stackrel{D}\rightarrow N(0,C^2\mu(A))$.
\end{cor}

\noindent
{\bf Proof:} At the beginning, we assume that there exists ${\bf t}\in\mathbb{R}^r$ and $V\subseteq\mathscr{B}({\cal L}^\top)$ such that $A=A_{{\bf t},V}$. If we partition $(0,t_1{\bf u}_1]\times\cdots\times(0,t_r{\bf u}_r]$ into countable small rectangles, denoted by ${\cal A}=\{U_i: i\in\mathbb{N}\}$, then we can express $A_{{\bf t},V}=\bigcup_{i=1}^\infty U_i\times V$. By theorem \ref{thm:functional central limit theorem strong mixing}, $M_{z,{\cal L}}({\cal A})\stackrel{D}\rightarrow N(0,D_{\cal A})$, where $D_{\cal A}$ is a diagonal matrix determined by the property of ${\cal A}$ and it satisfies all of the assumptions of $\sigma$-finite measures in ${\cal L}$. Therefore, there exists a $\sigma$-finite measure $\tilde\mu$ on ${\cal L}$ such that $M_{z,{\cal L}}(A_{{\bf t},V})\stackrel{D}\rightarrow N(0,\tilde\mu(A_{{\bf t},V}))$. Note that ${\cal A}_{V}$ is a $\pi$-system \cite[P. 42]{billingsley1995}, we conclude that $\tilde\mu$ can be uniquely determined. Then, there is $M_{z,L}(A)\stackrel{D}\rightarrow N(0,\tilde\mu(A))$ for any $A\in\mathscr{B}(\mathbb{R}^d)$. By the expression of $V[N(A_{z,L})]$ given by (\ref{eq:covariance of counts in two events}), we conclude that $\tilde\mu(A)$ is proportional to $\mu(A)$, implying the conclusion.  \qed

\bigskip
A main interest in practice is to estimate the first-order intensity function $\lambda({\bf s})$ under substationarity. As $\lambda({\bf s})$ only varies in ${\cal L}^\perp$, it is equivalent to estimate $\lambda({\bf s}_{{\cal L}^\perp})$ and ${\cal L}$ together. Since it is generally inappropriate to model $\lambda({\bf s}_{{\cal L}^\perp})$ parametrically, we propose a nonparametric way to estimate it. Note that ${\cal L}$ can be formulated by a rotation of a linear subspace spanned by coordinates, we propose a parametric way to estimate it. Therefore, we classify our estimation as a semiparametric approach. The functional central limit theorems given by Theorems \ref{thm:functional central limit theorem for Poisson substationary SPP} and \ref{thm:functional central limit theorem strong mixing} provide the theoretical basis of the approach. 

\section{Estimation}
\label{sec:estimation}

Let ${\cal N}$ be substationary in ${\cal L}\subseteq\mathbb{R}^d$. Assume points of ${\cal N}$ are only collected in bounded ${\cal S}\in\mathscr{B}(\mathbb{R}^d)$ such that they can be represented by ${\cal N}({\cal S})$. Our main interest is to estimate $\lambda({\bf s}_{{\cal L}^\perp})$ and ${\cal L}$ simultaneously by ${\cal N}({\cal S})$. Since ${\cal L}$ is unknown, we propose a two-step method to estimate them. In the first step, we estimate $\lambda({\bf s}_{{\cal L}^\perp})$ with a given ${\cal L}$, where a nonparametric way is adopted. In the second step, we estimate ${\cal L}$, where a parametric way is adopted. The second step needs the formulation of the estimator in the first step. 

We propose a kernel-based method to estimate $\lambda({\bf s})$ for a given ${\cal L}$. We investigate the usual kernel-based method without using substationarity \cite{diggle1985}. It provides an estimator of $\lambda({\bf s})$ as
\begin{equation}
\label{eq:general kernel estimator of the first-order}
\hat\lambda_{h}({\bf s})=C_{h}^{-1}({\bf s})\int_{\cal S} K_{h}({\bf s}'-{\bf s}) N(d{\bf s}'), 
\end{equation}
where $K_{h}({\bf s})=K({\bf s}/h)/h^d$ with bandwidth $h\in\mathbb{R}$ is a kernel density function on $\mathbb{R}^d$ and $C_{h}({\bf s})=\int_{\cal S}K_{h}({\bf s}'-{\bf s})d{\bf s}'$ is the Berman-Diggle boundary correction \cite{berman1989}. By Campbell's Theorem, we obtain 
\begin{equation}
\label{eq:expected value of the general kernerl estimator}
{\rm E}[\hat\lambda_{h}({\bf s})]
=C_h^{-1}({\bf s})\int_{\cal S}K_h({\bf s}'-{\bf s})\lambda({\bf s}')d{\bf s}'
\end{equation}
and
\begin{equation}
\label{eq:variance of the general kernerl estimator}
\eqalign{
{\rm V}[\hat\lambda_h({\bf s})]=&C_h^{-2}({\bf s})\int_{\cal S}\int_{\cal S}K_h({\bf s}'-{\bf s})K_h({\bf s}''-{\bf s})[g({\bf s}',{\bf s}'')-1]\lambda({\bf s}')\lambda({\bf s}'')d{\bf s}''d{\bf s}'\cr
&+C_h^{-2}({\bf s})\int_{\cal S}K_h^2({\bf s}'-{\bf s})\lambda({\bf s})d{\bf s}.
}\end{equation}

We modify (\ref{eq:general kernel estimator of the first-order}) for a substationary ${\cal N}$ in ${\cal L}$. We obtain an estimator of $\lambda({\bf s}_{{\cal L}^\perp})$ (or $\lambda({\bf s})$, equivalently) as
\begin{equation}
\label{eq:kernel estimator of the first-order substationarity}
\hat\lambda_{h,{\cal L}^\perp}({\bf s}_{{\cal L}^\perp})=C_{h,{\cal L}^\perp}^{-1}({\bf s}_{{\cal L}^\perp})\int_{\cal S} K_{h,{\cal L}^\perp}({\bf s}_{{\cal L}^\perp}'-{\bf s}_{{\cal L}^\perp}) N(d{\bf s}'),
\end{equation}
where $K_{h,{\cal L}^\perp}({\bf s}_{{\cal L}^\perp})=K({\bf s}_{{\cal L}^\perp}/h)/h^r$ with $h\in\mathbb{R}$ is a kernel density function on ${\cal L}^\perp$ and $C_{h,{\cal L}^\perp}({\bf s})=\int_{\cal S} K_{h,{\cal L}^\perp}({\bf s}_{{\cal L}^\perp}'-{\bf s}_{{\cal L}^\perp})d{\bf s}'$ is still the boundary correction. Still by Campbell's Theorem, we obtain 
\begin{equation}
\label{eq:expected value kernel estimator of the first-order substationarity}
{\rm E}[\hat\lambda_{h,{\cal L}^\perp}({\bf s}_{{\cal L}^\perp})]=C_{h,{\cal L}^\perp}^{-1}({\bf s}_{{\cal L}^\perp})\int_{\cal S} K_{h,{\cal L}^\perp}({\bf s}_{{\cal L}^\perp}'-{\bf s}_{{\cal L}^\perp})\lambda({\bf s}_{{\cal L}^\perp}')d{\bf s}',
\end{equation}
and
\begin{equation}
\label{eq:variance kernel estimator of the first-order substationarity}
\eqalign{
{\rm V}[\hat\lambda_{h,{\cal L}^\perp}({\bf s}_{{\cal L}^\perp})]=&C_{h,{\cal L}^\perp}^{-2}({\bf s}_{{\cal L}^\perp})\int_{\cal S}\int_{\cal S} K_{h,{\cal L}^\perp}({\bf s}_{{\cal L}^\perp}'-{\bf s}_{{\cal L}^\perp})K_{h,{\cal L}^\perp}({\bf s}_{{\cal L}^\perp}''-{\bf s}_{{\cal L}^\perp})[g({\bf s}',{\bf s}'')-1]\cr
&\lambda({\bf s}_{{\cal L}^\perp}')\lambda({\bf s}_{{\cal L}^\perp}'')d{\bf s}''d{\bf s}'+C_{h,{\cal L}^\perp}^{-2}({\bf s})\int_{\cal S}K_{h,{\cal L}^\perp}^2({\bf s}_{{\cal L}^\perp}'-{\bf s}_{{\cal L}^\perp})\lambda({\bf s}_{{\cal L}^\perp}')d{\bf s}'.\cr
}
\end{equation}
If $r=0$, then ${\cal L}=\{{\bf 0}\}$ and (\ref{eq:kernel estimator of the first-order substationarity}) becomes
\begin{equation}
\label{eq:estimator of intensity under stationarity}
\hat\lambda={n\over |{\cal S}|}.
\end{equation}
Since ${\cal N}$ is stationary in this case, the first-order intensity function is a constant, indicating that the estimator must be a constant. 

We compare the MSEs (mean square errors) of $\hat\lambda_h({\bf s})$ and $\hat\lambda_{h,{\cal L}^\perp}({\bf s}_{{\cal L}^\perp})$ as $z\rightarrow\infty$ in the case when ${\cal S}=A_{z,{\cal L}}$ for a bounded $A\in\mathscr{B}(\mathbb{R}^d)$. We find that the bias of $\hat\lambda_h({\bf s})$, which is given by ${\rm Bias}[\hat\lambda_h({\bf s})]={\rm E}[\hat\lambda_h({\bf s})]-\lambda({\bf s})$, can go to $0$ as $h\rightarrow0$, but it can simultaneously cause ${\rm V}[\hat\lambda_h({\bf s})]\rightarrow\infty$. To make ${\rm V}[\hat\lambda_h({\bf s})]$ small, we need to choose a large $h$, which increases the value of ${\rm Bias}[\hat\lambda_h({\bf s})]$. Thus, ${\rm MSE}[\hat\lambda_h({\bf s})]=\{{\rm E}[\hat\lambda_h({\bf s})]-\lambda({\bf s})\}^2+{\rm V}[\hat\lambda_h({\bf s})]$ cannot go to $0$ as $z\rightarrow\infty$. However, by a way to select $h$, we can make ${\rm MSE}[\hat\lambda_{h,{\cal L}^\perp}({\bf s}_{{\cal L}^\perp})]\rightarrow 0$ as $z\rightarrow\infty$. 

\begin{thm}
\label{thm:unbiasedness of estimators if h is small}
Let ${\cal N}$ be substationary in ${\cal L}$ and ${\cal S}=A_{z,{\cal L}}$ for a bounded $A\in\mathscr{B}(\mathbb{R}^d)$ with $|\partial A|=0$. Suppose all of conditions of Theorem \ref{thm:functional central limit theorem strong mixing} hold. Assume $\lambda({\bf s}_{{\cal L}^\perp})$ is positive and continuous in the interior of ${\cal S}$ and $\nu_r(A_{\bf v})$ is almost surely continuous in any ${\bf v}\in{\cal A}^\perp$. For an interior point ${\bf s}$ of $A$, if $h\rightarrow 0$ and $hz\rightarrow\infty$, then ${\rm MSE}[\hat\lambda_{h,{\cal L}^\perp}({\bf s}_{{\cal L}^\perp})]\rightarrow 0$ as $z\rightarrow\infty$.
\end{thm}

\noindent
{\bf Proof:} For an interior point of ${\bf s}\in A$, there is 
$$\eqalign{
{\rm E}[\hat\lambda_{h,{\cal L}^\perp}({\bf s}_{{\cal L}^\perp})]=&\left\{\int_{A_{z,{\cal L}}}K_{h,{\cal L}^\perp}({\bf s}_{{\cal L}^\perp}'-{\bf s}_{{\cal L}^\perp})d{\bf s}'\right\}^{-1}\int_{A_{z,{\cal L}}} K_{h,{\cal L}^\perp}({\bf s}_{{\cal L}^\perp}'-{\bf s}_{{\cal L}^\perp})\lambda({\bf s}_{{\cal L}^\perp}')d{\bf s}'\cr
=&\left\{\int_{A}K_{h,{\cal L}^\perp}({\bf s}_{{\cal L}^\perp}'-{\bf s}_{{\cal L}^\perp})d{\bf s}'\right\}^{-1}\int_A K_{h,{\cal L}^\perp}({\bf s}_{{\cal L}^\perp}'-{\bf s}_{{\cal L}^\perp})\lambda({\bf s}_{{\cal L}^\perp}')d{\bf s}'\cr
=&\left\{\int_{{\cal L}^\perp}\nu_r(A_{{\bf s}_{{\cal L}^\perp}+h{\bf v}})K({\bf v})d{\bf v}\right\}^{-1}\int_{{\cal L}^\perp}\nu_r(A_{{\bf s}_{{\cal L}^\perp}+h{\bf v}})K({\bf v})\lambda({\bf s}_{{\cal L}^\perp}+h{\bf v})d{\bf v}.
}$$
If $h\rightarrow 0$ as $z\rightarrow\infty$, then by the continuity of $\nu_r(A_{\bf v})$ and $\lambda({\bf s}_{{\cal L}^\perp})$ there is
$$\eqalign{
\lim_{z\rightarrow\infty}{\rm E}[\hat\lambda_{h,{\cal L}^\perp}({\bf s}_{{\cal L}^\perp})]
=&\lambda({\bf s}_{{\cal L}^\perp}).\cr
}$$  
By (\ref{eq:variance kernel estimator of the first-order substationarity}), there is 
$$\eqalign{
{\rm V}[\hat\lambda_{h,{\cal L}^\perp}({\bf s}_{{\cal L}^\perp})]=&\left\{\int_{A}K_{h,{\cal L}^\perp}({\bf s}_{{\cal L}^\perp}'-{\bf s}_{{\cal L}^\perp})d{\bf s}'\right\}^{-2}\int_{A}\int_{A} K_{h,{\cal L}^\perp}({\bf s}_{{\cal L}^\perp}'-{\bf s}_{{\cal L}^\perp})K_{h,{\cal L}^\perp}({\bf s}_{{\cal L}^\perp}''-{\bf s}_{{\cal L}^\perp})\cr
&\times\{g[{\bf s}',{\bf s}''+z({\bf s}_{\cal L}''-{\bf s}_{\cal L}')]-1\}\lambda({\bf s}_{{\cal L}^\perp}')\lambda({\bf s}_{{\cal L}^\perp}'')d{\bf s}''d{\bf s}'\cr
&+z^{-r}\left\{\int_{A}K_{h,{\cal L}^\perp}({\bf s}_{{\cal L}^\perp}'-{\bf s}_{{\cal L}^\perp})d{\bf s}'\right\}^{-2}\int_{A}K_{h,{\cal L}^\perp}^2({\bf s}_{{\cal L}^\perp}'-{\bf s}_{{\cal L}^\perp})\lambda({\bf s}_{{\cal L}^\perp}')d{\bf s}'.
}$$
By Theorem \ref{thm:functional central limit theorem strong mixing}, we conclude that the first term of the above goes to $0$ as $z\rightarrow\infty$. Therefore, we only need to study the second term. It is
$$\eqalign{
&{1\over h^rz^r}\left\{\int_{{\cal L}^\perp}\nu_r(A_{{\bf s}_{{\cal L}^\perp}+h{\bf v}})K({\bf v})d{\bf v}\right\}^{-2}\int_{{\cal L}^\perp}K^2({\bf v})\lambda({\bf s}_{{\cal L}^\perp}+h{\bf v})d{\bf v},
}$$
which goes to zero if $hz\rightarrow\infty$. \qed
 
\bigskip
{\it Example 1:} We interpret Theorem \ref{thm:unbiasedness of estimators if h is small} in a special case. Assume that ${\cal N}$ is substationary in $\mathbb{R}^2$ and ${\cal L}=\{(x,0):x\in\mathbb{R}\}$ such that $d=2$, $r=1$, and the first-order intensity function can be expressed as $\lambda({\bf s})=\lambda(y)$, where ${\bf s}=(x,y)$. Suppose ${\cal S}=[0,z]\times[0,\omega]$ such that observations of ${\cal N}$ can be expressed by points within $[0,z]\times[0,\omega]$, denoted by ${\bf s}_1,\cdots,{\bf s}_n$, where $n={N({\cal S})}$ is the total number of observed points. If we choose $K({\bf s})=(2\pi)^{-1}e^{-(x^2+y^2)/2}$ for the case when substationarity is not accounted for, then $K_h({\bf s})=\phi(x/h)\phi(y/h)/h^2=(2\pi h^2)^{-1}e^{-(x^2+y^2)/(2h^2)}$, where $\phi$ is the PDF of $N(0,1)$. By (\ref{eq:general kernel estimator of the first-order}), there is
$$\eqalign{
{\rm E}[\hat\lambda_h({\bf s})]=\left\{\int_0^z\int_0^\omega{1\over 2\pi h^2}e^{-{(x'-x)^2+(y'-y)^2\over 2h^2}}dy'dx'\right\}^{-1}\int_0^z\int_0^\omega{1\over 2\pi h^2}e^{-{(x'-x)^2+(y'-y)^2\over 2h^2}}\lambda({\bf s}')d{\bf s}'.
}$$
Then, $\lim_{h\rightarrow 0}{\rm E}[\hat\lambda_h({\bf s})]=\lambda({\bf s})$, implying that the bias of $\hat\lambda_h({\bf s})$ can only disappear as $h\rightarrow\infty$ but this can make ${\rm V}[\hat\lambda_h({\bf s})]$ large. If we choose $K(y)=(2\pi)^{-1/2}e^{-y^2/2}$ for the case when substationarity is accounted for, then $K_{h,{\cal L}^\perp}(y)=\phi(y/h)/h=(\sqrt{2\pi}h)^{-1}e^{-y^2/(2h^2)}$. By (\ref{eq:kernel estimator of the first-order substationarity}), there is
$${\rm E}[\hat\lambda_{h,{\cal L}^\perp}(y)]=\left\{\int_0^\omega {1\over\sqrt{2\pi}h}e^{-{(y'-y)^2\over 2h^2}}dy'\right\}^{-1}\int_0^\omega {1\over\sqrt{2\pi}h}e^{-{(y'-y)^2\over 2h^2}}\lambda(y)dy.$$
Then, $\lim_{h\rightarrow 0}{\rm E}[\hat\lambda_{h,{\cal S}^\perp}({\bf s})]=\lambda({\bf s})$, implying that the bias of $\hat\lambda_{h,{\cal L}^\perp}({\bf s})$ also disappears as $h\rightarrow\infty$. By (\ref{eq:variance kernel estimator of the first-order substationarity}), there is
$$\eqalign{
{\rm V}[\hat\lambda_{h,{\cal L}^\perp}(y)]=&\left\{\int_0^\omega {1\over\sqrt{2\pi}h}e^{-{(y'-y)^2\over 2h^2}}dy'\right\}^{-2}\int_0^\omega\int_0^\omega {1\over 2\pi h^2}e^{-{(y'-y)^2+(y''-y)^2\over 2h^2}}\lambda(y')\lambda(y'')\cr
&\times\left\{{1\over z}\int_0^z \{g[(0,y'),(x'',y'')]-1\}dx''\right\}dy'dy''\cr
&+{1\over z}\left\{\int_0^\omega {1\over\sqrt{2\pi}h}e^{-{(y'-y)^2\over 2h^2}}dy'\right\}^{-2}\int_0^\omega {1\over\sqrt{2\pi}h}e^{-{(y'-y)^2\over 2h^2}}\lambda(y')dy'.
}$$
If all conditions of Theorem \ref{thm:functional central limit theorem strong mixing} hod, then $\lim_{x''\rightarrow\infty}g[(0,y'),(x'',y'')]-1=0$. Thus, the first term of above goes to $0$ as $z\rightarrow\infty$. Further, we conclude the second term goes to zero if $zh\rightarrow\infty$. Thus, we have the conclusion of Theorem \ref{thm:unbiasedness of estimators if h is small}. 

\bigskip
As ${\cal L}$ is also unknown, we should have a way to estimate ${\cal L}$ in the usage of $\hat\lambda_{h,{\cal L}^\perp}({\bf s})$. Let ${\cal L}={\rm span}\{{\bf u}_1,\cdots,{\bf u}_r\}$, where ${\bf u}_1,\cdots,{\bf u}_r$ are orthonormal vectors of ${\cal L}$. Then, it is enough to provide an estimator of $\{{\bf u}_1,\cdots,{\bf u}_r\}$ is our method. If $r=0$, then ${\cal N}$ is not substationary in any linear subspace of $\mathbb{R}^d$. If $r=d$, then ${\cal N}$ is stationary in the entire $\mathbb{R}^d$. Otherwise, ${\cal N}$ is substationary in ${\cal L}$ but nonstationary in $\mathbb{R}^d$. Note that ${\cal L}$ can be represented by an orthogonal projection ${\bf Q}$ in $\mathbb{R}^d$. Let $\mathscr{Q}$ be the collection of the orthogonal projections from $\mathbb{R}^d$ to an $r$-dimensional linear subspace. Estimation of ${\cal L}$ is equivalent to estimation of ${\bf Q}\in\mathscr{Q}$. Let
\begin{equation}
\label{eq:likelihood function}
\ell[\lambda({\bf s})]=\sum_{i=1}^n \log\lambda({\bf s})-\int_{\cal S}\lambda({\bf s})d{\bf s}
\end{equation}
be the loglikelihood function of ${\cal N}({\cal S})$ if ${\cal N}$ is Poisson. Then, $\ell[\lambda({\bf s})]$ can be treated as the composite loglikelihood of ${\cal N}({\cal S})$ if ${\cal N}$ is non-Poisson \cite{guan2010}. Therefore, we can estimate ${\bf Q}$ by 
\begin{equation}
\label{eq:estimator of the space}
\hat{\bf Q}_h=\mathop{\arg\!\max}_{{\bf Q}\in\mathscr{Q}}\ell[\hat\lambda_{h,{\cal L}^\perp}({\bf s})].
\end{equation} 
 To apply (\ref{eq:estimator of the space}), we need to provide a way to determine the best $h$ in $\hat{\bf Q}_h$, where we recommend using the generalized cross validation (GCV) approach \cite{golub1979}.

\section{Simulation}
\label{sec:simulation}

We carried out a simulation study to evaluate the performance of $\hat\lambda_{h,{\cal L}^\perp}({\bf s})=\hat\lambda_{h,{\cal L}^\perp}({\bf s}_{{\cal L}^\perp})$ given by (\ref{eq:kernel estimator of the first-order substationarity}). We simulated realizations from Poisson and Poisson cluster SPPs in a rectangle region ${\cal S}=[0,z]\times[0,\omega]$, the region used in Example 1. We chose $\omega=1$ in our simulation. We selected these processes because they are popular in modeling ecological, environmental, geographical data. In both processes, we chose the first-order intensity function as
\begin{equation}
\label{eq:first-order intensity function in simulation}
\lambda({\bf s})={100\Gamma^2(a)\over\Gamma(2a)}y^{a-1}(1-y)^{a-1}
\end{equation}
for a selected $a\ge 1$ such that we always had $\kappa={\rm E}[N({\cal S})]=100z$. Note that $\lambda(s)/100$ is the PDF of $Beta(a,a)$ distribution. We chose $a=1.0,1.5,2.0,2.5,3.0$ in our simulations. If $a=1$, then ${\cal N}$ was stationary in the entire $\mathbb{R}^2$; otherwise, it was only substationary in ${\cal L}=\{(x,0): x\in\mathbb{R}\}$. Since ${\cal L}$ might be unknown, we also evaluated the performance of $\hat{\cal L}$, the estimator of ${\cal L}$ given by (\ref{eq:estimator of the space}).

To obtain a Poisson SPP, we first generated the number of points from the $Poisson(\kappa)$ distribution and then identically and independently generated the locations of these points. The horizontal values of these points were generated from the uniform distribution on $[0,z]$. The vertical values of these points were generated from the $Beta(a,a)$ distribution. To obtain a Poisson cluster SPP,  we first generated their parent points from a Poisson SPP with its first-order intensify function equal to $\lambda({\bf s})/\gamma$ by the same method for the Poisson SPP. After parent points were derived, we generated offspring points. Each parent point generated $Poisson(\gamma)$ offspring points independently. The position of each offspring point relative to its parent point was defined as a radially symmetric Gaussian random variable with a standard deviation $\sigma$. We chose $\gamma=5$ and $\sigma=0.02$ in all the cases of Poisson cluster SPPs that we studied. 

We studied two cases in the implementation of $\hat\lambda_{h,{\cal L}}({\bf s})$. In the first case, we assumed that ${\cal L}$ was known such that we could directly apply (\ref{eq:kernel estimator of the first-order substationarity}). We chose $K_{h,{\cal L}^\perp}(y)=\phi(y/h)/h$ as the density of $N(0,h^2)$. Then, we had $C_{h,{\cal L}^\perp}(y)=z\{\Phi[(\omega-y)/h]-\Phi(-y/h)\}$, where $\Phi$ is the CDF of $N(0,1)$, indicating that
\begin{equation}
\label{eq:estimator known L simulation}
\hat\lambda_{h,{\cal L}^\perp}(y)=\left\{z\left[\Phi({\omega-y\over h})-\Phi(-{y\over h})\right]\right\}^{-1}\sum_{i=1}^n {1\over\sqrt{2\pi}h}e^{-{(y_i-y)^2\over 2h^2}}, 0<y<\omega.
\end{equation}
In the second case, we assumed that ${\cal L}$ was unknown. We also needed to estimate ${\cal L}$. Note that any one-dimensional linear subspace of $\mathbb{R}^2$ can be expressed as
\begin{equation}
\label{eq:linear subspace with angle}
{\cal L}_{\theta}=\{(u\cos\theta,u\sin\theta):u\in\mathbb{R}\}, \theta\in[-{\pi\over 2},{\pi\over 2}),
\end{equation}
indicating that its vertical space is
\begin{equation}
\label{eq:vertical linear subspace with angle}
{\cal L}_{\theta}^\perp=\{(-v\sin\theta,v\cos\theta):v\in\mathbb{R}\}, \theta\in[-{\pi\over 2},{\pi\over 2}).
\end{equation}
We chose $K_{h,{\cal L}_\theta^\perp}(v)=\phi(v/h)/h$ on ${\cal L}_{\theta}^\perp$. 

To apply (\ref{eq:kernel estimator of the first-order substationarity}), we computed the analytic expression of $C_{h,{\cal L}_\theta^\perp}(v)$. If $\theta=0$, then 
$$C_{h,{\cal L}_\theta^\perp}(v)=z\left[\Phi({\omega-v\over h})-\Phi(-{v\over h})\right].$$
If $\theta=-\pi/2$, then 
$$C_{h,{\cal L}_\theta^\perp}(v)=\omega\left[\Phi({z-v\over h})-\Phi(-{v\over h})\right].$$
If $0<\theta<\pi/2$, then 
$$\eqalign{
C_{h,{\cal L}_\theta^\perp}(v)=&\left({z\over\cos\theta}+{v\over\sin\theta\cos\theta}\right)\left\{\Phi\left({(\omega\cos\theta-z\sin\theta)\wedge 0-v\over h}\right)-\Phi\left({-z\sin\theta-v\over h} \right)\right\}\cr
&+{h\over\sin\theta\cos\theta}\left[\phi\left({-z\sin\theta-v\over h}\right)-\phi\left({(\omega\cos\theta-z\sin\theta)\wedge 0-v\over h}\right)\right]\cr
&+\left({z\over\cos\theta}\wedge{\omega\over\sin\theta}\right)\left\{\Phi\left({(\omega\cos\theta-z\sin\theta)\vee 0-v\over h}\right)-\Phi\left({(\omega\cos\theta-z\sin\theta)\wedge 0-v\over h}\right)\right\}\cr
&+\left({\omega\cos\theta-v\over\sin\theta\cos\theta}\right)\left\{\Phi\left({\omega\cos\theta-v\over h}\right)-\Phi\left[{(\omega\cos\theta-z\sin\theta)\vee 0-v\over h}\right]\right\}\cr
&-{h\over\sin\theta\cos\theta}\left\{\phi\left({(\omega\cos\theta-z\sin\theta)\vee 0-v\over h}\right)-\phi\left({\omega\cos\theta-v\over h}\right) \right\}.\cr
},$$
where $-z\sin\theta\le v\le \cos\theta$. If $-\pi/2<\theta<0$, then 
$$\eqalign{
C_{h,{\cal L}_\theta^\perp}(v)=&-{v\over\sin\theta\cos\theta}\left\{\Phi\left({(-z\sin\theta)\wedge(\omega\cos\theta)-v\over h}\right)-\Phi\left(-{v\over h}\right)\right\}\cr
&-{h\over\sin\theta\cos\theta}\left\{\phi\left(-{v\over h}\right)-\phi\left({(-z\sin\theta)\wedge(\omega\cos\theta)-v\over h}\right)\right\}\cr
&+\left[{z\over\cos\theta}\wedge\left(-{\omega\over\sin\theta}\right)\right]\left\{\Phi\left({(-z\sin\theta)\vee(\omega\cos\theta)-v\over h}\right)-\Phi\left({(-z\sin\theta)\wedge(\omega\cos\theta)-v\over h}\right)\right\}\cr
&+{z\sin\theta-\omega\cos\theta+v\over\sin\theta\cos\theta}\left\{\Phi\left({-z\sin\theta+\omega\cos\theta-v\over h}\right)-\Phi\left({(-z\sin\theta)\vee(\omega\cos\theta)-v\over h}\right)\right\}\cr
&+{h\over\sin\theta\cos\theta}\left\{\phi\left({(-z\sin\theta)\vee(\omega\cos\theta)-v\over h}\right)-\phi\left({-z\sin\theta+\omega\cos\theta-v\over h}\right)\right\},\cr
}$$
where $0\le v\le -z\sin\theta+\cos\theta$. For a given $\theta\in[-\pi/2,\pi/2)$, we calculated $\hat\lambda_{h,{\cal L}_\theta^\perp}({\bf s}_{{\cal L}_\theta^\perp})$ by (\ref{eq:kernel estimator of the first-order substationarity}) as
\begin{equation}
\label{eq:estimate linear subspace with angle}
\hat\lambda_{h,{\cal L}_\theta^\perp}(v)=C_{h,{\cal L}_\theta^\perp}^{-1}(v)\sum_{i=1}^n {1\over\sqrt{2\pi}h}e^{-{(y_i\cos\theta -x_i\sin\theta-v)^2\over 2h^2}}
\end{equation}
for $(-z\sin\theta)\wedge 0\le v\le \cos\theta+(-z\sin\theta)\vee0$, where points were given by ${\bf s}_i=(x_i,y_i)$ for $i=1,\cdots,n$. We calculated $\hat\theta_h$ by (\ref{eq:estimator of the space}) and (\ref{eq:estimate linear subspace with angle}). We defined $\mathscr{Q}=\{\theta:{\bf Q}_{\theta}\}$ in the implementation of (\ref{eq:estimator of the space}), where ${\bf Q}_{\theta}{\bf s}=y\cos\theta-x\sin\theta$ was an orthogonal project from $\mathbb{R}^2$ to ${\cal L}_{\theta}$. The estimator $\hat\theta_h$ was the value of $\theta$ corresponding to $\hat{\bf Q}_h$ given by (\ref{eq:estimator of the space}). With $\hat\theta_h$, we calculated the value of $\hat\lambda_{h,\hat{\cal L}^\perp}(v)$ with $\hat{\cal L}={\cal L}_{\hat\theta_h}$, which was treated as the estimator of $\lambda({\bf s})$ under substationarity with an unknown ${\cal L}$. It was compared with $\hat\lambda_{h,{\cal L}^\perp}(y)$, the estimator of $\lambda({\bf s})$ with a known ${\cal L}$.

We evaluated the performance of the MSE (mean squares error) of $\hat\theta_h$ and the MISE (mean integrated square error) of $\hat\lambda_{h,{\cal L}_\theta^\perp}(v)$ for selected $a$, $z$, and $h$. The performance of $\hat\lambda_{h,{\cal L}_\theta^\perp}(v)$ was compared with that of $\hat\lambda_h({\bf s})$ given by (\ref{eq:general kernel estimator of the first-order}) and $\hat\lambda$ given by (\ref{eq:estimator of intensity under stationarity}), where we chose $K({\bf s})$ as the density of the standard bivariate normal distribution in the computation of $\hat\lambda_h({\bf s})$.

\begin{table}
\caption{\label{tab:performance of estimator of theta} Simulations (with 1,000 replications) for root MSEs of $\hat\theta_h$ (given by degrees) with respect to selected $a$, $z$, and $h$ in the Poisson and Poisson cluster processes.}
\begin{center}
\begin{tabular}{cccccccccc}\hline
 & & \multicolumn{4}{c}{$h$ for Poisson} & \multicolumn{4}{c}{$h$ for Poisson Cluster}\\
$a$ & $z$ & $0.01$ & $0.02$ & $0.05$ & $0.1$ & $0.01$ & $0.02$ & $0.05$ & $0.1$ \\\hline 
$1.5$&$1$&$4.24$&$4.14$&$4.61$&$5.12$&$4.51$&$4.52$&$4.84$&$5.22$\\
&$2$&$4.18$&$3.87$&$3.85$&$3.23$&$4.07$&$4.19$&$4.77$&$4.73$\\
&$5$&$2.84$&$2.22$&$1.12$&$1.03$&$4.51$&$4.13$&$3.64$&$3.06$\\
&$10$&$1.55$&$0.45$&$0.32$&$0.32$&$4.83$&$4.25$&$3.07$&$1.59$\\
$2.0$&$1$&$3.75$&$4.28$&$4.66$&$4.54$&$4.24$&$4.50$&$4.92$&$5.04$\\
&$2$&$3.39$&$2.83$&$2.63$&$2.58$&$3.82$&$3.90$&$4.07$&$4.41$\\
&$5$&$1.27$&$0.82$&$0.56$&$0.50$&$3.20$&$2.98$&$2.06$&$1.73$\\
&$10$&$0.32$&$0.20$&$0.19$&$0.20$&$1.81$&$1.67$&$0.58$&$0.72$\\
$2.5$&$1$&$3.72$&$4.03$&$4.07$&$4.10$&$4.03$&$4.49$&$4.43$&$4.90$\\
&$2$&$2.78$&$2.78$&$2.05$&$1.93$&$3.73$&$3.90$&$3.66$&$3.90$\\
&$5$&$0.77$&$0.54$&$0.43$&$0.39$&$2.18$&$1.94$&$1.22$&$1.25$\\
&$10$&$0.22$&$0.19$&$0.13$&$0.16$&$0.89$&$0.62$&$0.39$&$0.42$\\
$3.0$&$1$&$3.78$&$4.04$&$4.05$&$3.73$&$4.08$&$4.30$&$4.84$&$4.86$\\
&$2$&$2.97$&$2.45$&$1.70$&$1.62$&$3.56$&$3.54$&$3.79$&$3.47$\\
&$5$&$0.69$&$0.48$&$0.37$&$0.37$&$1.57$&$1.52$&$0.92$&$0.88$\\
&$10$&$0.24$&$0.15$&$0.11$&$0.14$&$0.48$&$0.49$&$0.34$&$0.35$\\\hline
\end{tabular}
\end{center}
\end{table}

We simulated $1000$ realizations for each selected cases. To evaluate the performance of $\hat\theta_h$, we computed its MSE value by $\sum_{i=1}^{1000}\hat\theta_{hi}^2/1000$, where $\hat\theta_{hi}$ was the value of $\hat\theta_h$ in the $i$th realization (Table \ref{tab:performance of estimator of theta}). We did not put the case when $a=1$ in the table as $\theta$ was not well-defined. The results showed that the root MSEs of $\hat\theta_h$ were all close to $0$, indicating that the estimator was accurate. The MSEs of $\hat\theta_h$ decreased as $z$ increased. This was interpreted by Theorem \ref{thm:unbiasedness of estimators if h is small}. The MSEs decreased as $a$ increased since the strength of nonstationarity increased as $a$ became large. For the same $a$ and $z$ values, the MSEs of $\hat\theta_h$ was also affected by the bandwidth $h$ in the kernel approach is always an important issue to be investigated. In all the cases that we studied, the MSEs of $\hat\theta_h$ in the Poisson SPPs was always lower than those in the Poisson cluster SPPs. This was expected as for the same $\kappa$ value the number of independent clusters in the Poisson cluster SPPs was lower than the number of independent points in the Poisson SPPs. 

\begin{table}
\caption{\label{tab:performance of estimator of intensity function}Simulations (with 1,000 replications) of root MSEs of $\hat\lambda_{h,{\cal L}^\perp}(y)$, $\hat\lambda_{h,{\cal L}_{\hat\theta_h}^\perp}(v)$, $\hat\lambda_h({\bf s})$, and $\hat\lambda$ with respected to selected $a$, $z$, and $h$ in the Poisson and Poisson cluster processes.}
\begin{center}
\begin{tabular}{ccccccccccc}\hline
& & & \multicolumn{4}{c}{Poisson} & \multicolumn{4}{c}{Poisson Cluster}\\
$a$ & $z$ & $h$ & $\hat\lambda_{h,{\cal L}^\perp}(y)$ & $\hat\lambda_{h,\hat{\cal L}^\perp}(v)$ & $\hat\lambda_h({\bf s})$ & $\hat\lambda$ & $\hat\lambda_{h,{\cal L}^\perp}(y)$ & $\hat\lambda_{h,\hat{\cal L}^\perp}(v)$ & $\hat\lambda_h({\bf s})$ & $\hat\lambda$ \\\hline
$1$&$1$&$0.05$&$25.13$&$36.95$&$61.17$&$10.33$&$54.43$&$60.22$&$133.01$&$23.24$\\
   &   &$0.10$&$17.31$&$17.95$&$32.19$&$9.20$&$39.23$&$40.85$&$70.90$&$21.27$\\
   &$2$&$0.05$&$17.48$&$19.45$&$59.72$&$6.55$&$39.44$&$45.13$&$132.72$&$16.81$\\
   &   &$0.10$&$13.16$&$13.95$&$31.60$&$8.03$&$29.17$&$31.69$&$70.90$&$16.57$\\
   &$5$&$0.05$&$11.21$&$13.56$&$59.10$&$4.87$&$24.21$&$31.22$&$130.30$&$9.50$\\
   &   &$0.10$&$8.25$&$9.82$&$31.17$&$4.83$&$18.45$&$22.35$&$68.69$&$10.81$\\
   &$10$&$0.05$&$7.45$&$10.96$&$58.60$&$2.49$&$16.32$&$25.29$&$129.79$&$6.13$\\
   &   &$0.10$&$5.64$&$7.91$&$30.42$&$3.26$&$13.29$&$18.42$&$67.95$&$7.64$\\
$2$&$1$&$0.05$&$24.08$&$26.00$&$59.14$&$45.91$&$54.52$&$57.75$&$130.70$&$50.25$\\
   &   &$0.10$&$21.31$&$21.90$&$33.93$&$45.66$&$40.20$&$41.14$&$70.03$&$49.43$\\
   &$2$&$0.05$&$17.58$&$19.94$&$58.17$&$45.18$&$38.16$&$43.02$&$128.43$&$47.38$\\
   &   &$0.10$&$17.23$&$19.44$&$32.98$&$45.25$&$28.95$&$32.67$&$69.97$&$47.12$\\
   &$5$&$0.05$&$12.00$&$12.76$&$57.96$&$44.95$&$24.67$&$29.93$&$127.09$&$45.75$\\
   &   &$0.10$&$14.40$&$15.67$&$32.44$&$44.91$&$21.59$&$24.55$&$67.69$&$45.96$\\
   &$10$&$0.05$&$8.92$&$9.33$&$57.41$&$44.82$&$17.15$&$19.44$&$126.20$&$45.24$\\
   &   &$0.10$&$13.64$&$14.12$&$32.28$&$44.83$&$16.93$&$18.46$&$66.67$&$45.21$\\
$3$&$1$&$0.05$&$23.95$&$26.08$&$59.25$&$66.17$&$50.81$&$55.22$&$129.91$&$68.65$\\
   &   &$0.10$&$21.88$&$22.67$&$34.36$&$66.12$&$41.83$&$43.57$&$71.43$&$69.77$\\
   &$2$&$0.05$&$17.23$&$18.96$&$58.47$&$65.83$&$37.31$&$42.61$&$126.97$&$67.28$\\
   &   &$0.10$&$19.25$&$21.20$&$33.38$&$66.04$&$31.94$&$34.50$&$68.15$&$67.71$\\
   &$5$&$0.05$&$11.32$&$12.00$&$56.88$&$65.61$&$24.30$&$27.19$&$125.90$&$66.37$\\
   &   &$0.10$&$15.21$&$15.92$&$32.49$&$65.61$&$20.87$&$22.39$&$66.81$&$66.07$\\
   &$10$&$0.05$&$7.97$&$8.27$&$56.77$&$65.53$&$16.56$&$18.32$&$124.36$&$65.85$\\
   &   &$0.10$&$14.71$&$15.01$&$32.22$&$65.55$&$18.67$&$19.71$&$66.36$&$65.89$\\\hline
\end{tabular}
\end{center}
\end{table}

We also evaluated the performance of four different estimators of the first-order intensity functions. Although we studied all of the selected cases in our simulations, we only put some of them in Table \ref{tab:performance of estimator of intensity function} to reduce the size of the table. We used $\hat\lambda_{h,{\cal L}^\perp}(y)$ to represent the case when $\theta$ was known. We used $\hat\theta_{h,\hat{\cal L}^\perp}(v)$ to represented the case when $\theta$ was unknown. We used $\hat\lambda_h({\bf s})$ to represent the case when substationarity was not taken into account. We used $\hat\lambda$ to represent the case when stationarity was assumed. All of the minimum MSEs were reached by $\hat\lambda$ when $a=1$ as the SPPs were stationary in this case. The MSEs of $\hat\lambda$ increased in $a$ since the strength of nonstationarity became large as $a$ increased. For the same $a$ and $h$ values, the MSEs of $\hat\lambda_{h,{\cal L}^\perp}(y)$ and $\hat\lambda_{h,\hat{\cal L}^\perp}(v)$ decreased in $z$. We interpreted this by Theorem \ref{thm:unbiasedness of estimators if h is small}. The MSEs of $\hat\lambda_h({\bf s})$ did not vary significantly as $z$ changed since the size of the region was not a critical issue in its computation. For all of the cases with $a>1$ that we studied, the MSEs of $\hat\lambda_{h,{\cal L}^\perp}(y)$ and $\hat\lambda_{h,\hat{\cal L}^\perp}(v)$ were lower than those of $\hat\lambda_{h}({\bf s})$ and $\hat\lambda$, indicating that efficiency was gained by accounting for  substationarity. 

\section{Application}
\label{sec:application}

\begin{figure}
\centerline{\includegraphics[angle=270,width=100mm]{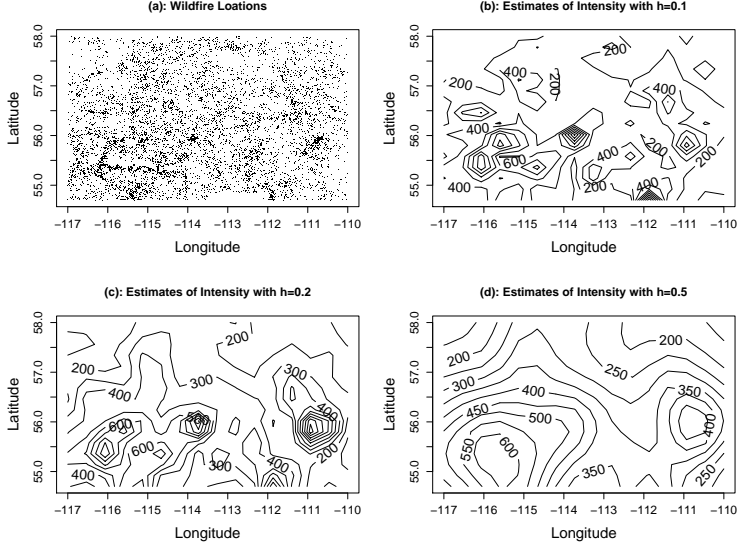}}
\caption{\label{fig:fire locations in a rectangle region} Wildfires locations and estimates of intensity under nonstationarity in Alberta Forests from 1996 to 2010 in the selected region, where bandwidths were given by degrees.}
\end{figure}

We applied our approach to the {\it Alberta Forest Wildfire} data. The {\it Alberta Forest Wildfire} data consisted of forest wildfire activities occurred in Alberta, Canada, from 1931 to 2012. The Canadian Alberta Forest Service initiated the modern era of wildfire record keeping in 1931. Since 1996, paper-based wildfire information was no long retained. The wildfire historical data were entered at the field level on the Fire Information Resource Evaluation System (FIREs), which can be freely downloaded from the internet. We collected the historical forest wildfire data from 1996 to 2010 within a rectangle spanned from $117$ longitude West to $110$ longitude West in the horizontal direction and from $54.7$ latitude North to $58$ latitude North in the vertical direction (Figure \ref{fig:fire locations in a rectangle region}(a)). We treated the rectangle as the study region in our approach. The region contained $8125$ wildfire occurrences with all of the three greatest wildfires occurred in Alberta forests during the $15$ years period. The greatest wildfire occurred in 2002 at $111.8$ longitude West and $55.5$ latitude North with area burned $2388.67{\rm km}^2$. The second greatest wildfire occurred in $1998$ at $116.5$ longitude West and $54.7$ latitude North with area burned $1631.38{\rm km}^2$. The third greatest wildfire occurred in $1998$ at $114.3$ longitude West and $47.5$ latitude West with area burned $1554.5{\rm km}^2$. The total burned area in the region was over $60\%$ of the total burned area in the entire region.   

The study region contained a large portion of boreal forests in Alberta, which was dominated in plain areas. A small portion of boreal forests of Alberta was in the mountain areas, located in the southwestern region of Alberta. We focused our study on the plain areas since tree densities and topographic conditions were significantly different between the mountain and plain areas. 

The geographical distribution of boreal forest wildfires is considered as a major dominant disturbance in the high latitude area of the North Hemisphere \cite{podur2002}. It has been pointed out that wildfire activities in boreal forest are significantly affected by latitude but not by longitude \cite{xiao2007}. It is expected to have low numbers of wildfire occurrences with high values of area burned in the north than those in the south \cite{zhang2014a}, indicating that substationarity might be assumed along the longitude. To confirm this, we calculated the estimates of $\lambda({\bf s})$ with the standard bivariate normal kernel via (\ref{eq:general kernel estimator of the first-order}) under nonstationarity. We used a few bandwidth values and found the results were not stable (Figure \ref{fig:fire locations in a rectangle region}(b), 2(c), and 2(d)). However, all of our results showed that the estimates of the intensity were high in the south but low in the north. 

We assumed fire occurrences were substationary in a linear space of $\mathbb{R}^d$, where the linear space was ${\cal L}={\cal L}_{\theta}$ given by (\ref{eq:linear subspace with angle}). We calculated $\hat\theta_h$ with a normal kernel in (\ref{eq:estimator of the space}). We treated $\hat\theta_h$ as an estimator of $\theta$ for a given $h$.  We compared values of $\hat\theta_h$ with various choices of $h$. We found that $\hat\theta_h$ was reliable. For instance, we got $\hat\theta_h=-0.002$ (given by arc degree, same as the following) if $h=0.01$, $\hat\theta_h=-0.001$ if $h=0.02$, $\hat\theta_h=-0.003$ if $h=0.05$, and $\hat\theta_h=-0.007$ if $h=0.1$. Therefore, we had $\hat\theta_h\approx0$, indicating that we might simply choose ${\cal L}={\cal L}_0$ in our estimation. To investigate this issue, we compared the values of $\ell[\hat\lambda_{h,\hat{\cal L}^\perp}({\bf s})]$ and $\ell[\hat\lambda_{h,{\cal L}_0^\perp}({\bf s})]$ with selected $h$ in (\ref{eq:likelihood function}). The values of $\ell[\hat\lambda_{h,\hat{\cal L}^\perp}({\bf s})]-\ell[\hat\lambda_{h,{\cal L}_0^\perp}({\bf s})]$ were $1.66$, $0.89$, $3.38$, and $5.3$ when $h$ were $0.01$, $0.02$, $0.05$, and $0.1$, respectively. Comparing these values with the differences of loglikelihood functions affected by $h$, which were often more than a few hundred, we concluded that the values of $\ell[\hat\lambda_{h,\hat{\cal L}^\perp}({\bf s})]-\ell[\hat\lambda_{h,{\cal L}_0^\perp}({\bf s})]$ could be ignored. Therefore, we could use $\theta=0$ in the computation of the estimates of the first-order intensity function.

\begin{figure}
\centerline{\includegraphics[angle=270,width=100mm]{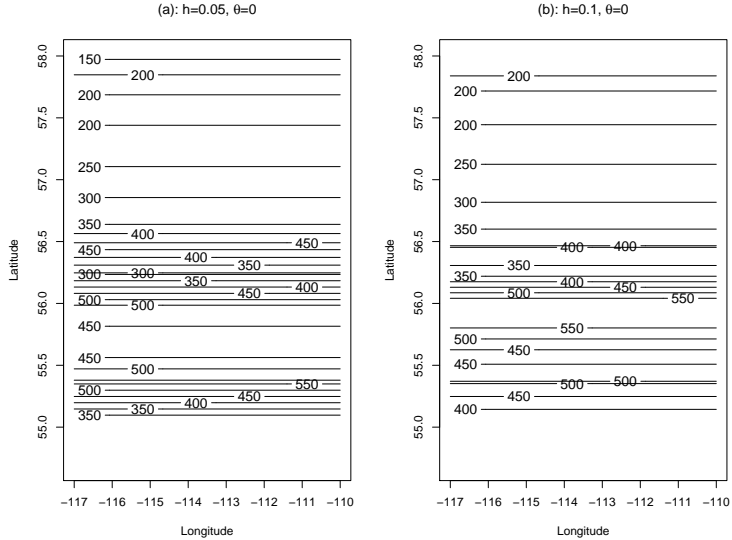}}
\caption{\label{fig:fire intensity under substationarity} Estimates of the first-order intensity in the {\it Alberta Forest Wildfire} data under substationarity along the longitude.}
\end{figure}

Simply using $\theta=0$, we obtained ${\cal L}_0=\{(x,0):x\in\mathbb{R}\}$. We used ${\cal L}_0$ to estimate the first-order intensity function of wildfire occurrences under substationarity. We computed values of $\hat\lambda_{h,{\cal L}_0^\perp}(y)$ with various choices of $h$. All of the results were close (e.g. as those displayed by Figure \ref{fig:fire intensity under substationarity}), indicating that our approach was reliable. We found that the intensity of wildfire occurrences was almost maximized at $55.8$ latitude North. It decreased fast to the north but slowly to the south. The north part was consistent with our previous conclusion but the south part was a concern. We studied the reason by looking at the terrestrial ecozones. We found that ecozones in the south of the study region was dominated by grassland, which might affect the occurrences of forest wildfires \cite{pitman2007,xiao2007}. 

\section{Discussion}
\label{sec:discussion}

In this article, we propose the concept of substationarity and provide a semiparametric method to estimate the first-order intensity function of a spatial point process. The method is modified from the classical kernel density estimation for random variables. Classical kernel density estimation is formulated under the assumption that sampling data are collected identically and independently from a continuous distribution. This assumption is violated because the dependence structure is often present in spatial point data. A common way to account for dependence structures in SPPs is to use the second-order intensity functions. As specific relationship between the first-order and the second-order intensity functions can be formulated under the concept of SOIRs, it is possible to have methods to account for both the first-order and the second-order intensity functions simultaneously under the concept of SOIRs. 

Although we have only discussed the kernel-based approach, two other nonparametric or semiparmetric approaches may also be considered. The {\it local polynomial} approach is modified from the kernel approach \cite{cleveland1988,fan1993}. It is based on the idea of the weighted localized polynomial regression, where the weights are determined by kernel functions of explanatory variables. The {\it smoothing spline} approach estimates a smooth function by minimizing a penalized likelihood function \cite{gu2013,kimeldorf1971,wahba1990}. The penalized likelihood function has two terms. The negative loglikelihood term controls the goodness-of-fit value. The penalty term controls the smoothness value. Both the local polynomial and the smooth spline approaches can be used to estimate the intensity functions of SPPs under substationarity. 

As a relative concept, nonsubstationarity is also an important concept for spatial point data. A nonsubstationarity approach must be adopted if assumptions of substationarity are violated. Based on the concept of substationarity, a few possible ways may be proposed for nonsubstationarity. An easy way is to borrow the idea of additive models in nonparametric statistics \cite{friedman1981,hastie1990}. Assume intensity functions of a nonsubstationary SPP can be expressed by the sum of intensity functions of a few substationary SPPs. If the linear space of the substationary SPPs are different such that their intersection only contains the origin, then the additive model provides nonsubstatioary intensity functions. The structure of additive models for nonsubstationarity in SPPs is essentially different from the structure of additive models in nonparametric statistics. Additive models in SPPs attempt to model additivity by intensity functions. Additive model in nonparametric statistics attempt to model additivity by mean structures. Additive models in SPPs contain dependence structures but additive models in nonparametric statistics do not. This is an interesting research question to be studied in the future.


\begin{thebibliography}{}
\bibitem{baddeley2000}
Baddeley, A.J., M\o ller, J. and Waagepetersen, R. (2000). Non- and semi-parametric estimation of interaction in inhomogeneous point patterns. {\it Statistica Neerlandica}, {\bf 54}, 329-350.
\bibitem{benes2005}
Ben\v es, V., Bodl\'ak, K., M\o ller, J., and Waagepetersen, R. (2005). A case study on point process modelling in disease mapping. {\it Image Analysis and Stereology}, {\bf 24}, 159-168.
\bibitem{berman1989}
Berman, M. and Diggle, P.J. (1989). Estimating weighted integrals of the second-order intensity of a spatial point process. {\it Journal of the Royal Statistical Society Series B}, {\bf 51}, 81-92.
\bibitem{besag1977}
Besag, J. (1977). Contribution to the discussion of Dr. Ripley's paper. {\it Journal of the Royal Statistical Society Series B}, {\bf 39}, 193-195.
\bibitem{billingsley1995}
Billingsley, P. (1995). {\it Probability and Measure}. Wiley, New York.
\bibitem{cleveland1988}
Cleveland, W.S. and Devlin, S. (1988). Locally weighted regression: an approach to regression analysis by local fitting. {\it Journal of the American Statistical Association}, {\bf 83}, 596-610.
\bibitem{diggle1985}
Diggle, P.J. (1985). A kernel method for smoothing point process data. {\it Applied Statistics}, {\bf 34}, 138-147.
\bibitem{diggle2003}
Diggle, P.J. (2003). {\it Statistical Analysis of Spatial Point Patterns, 2nd Edition}, New York: Wiley.
\bibitem{diggle2006}
Diggle, P.J. (2006). Spatio-temporal point processes, partial likelihood, foot and mouth disease. {\it Statistical Methods in Medical Research}, {\bf 16}, 325-336.
\bibitem{diggle2007}
Diggle, P., Rubio, G., Brown, P.E., Chetwynd, A.G., and Gooding, S. (2007). Second-order analysis of inhomogeneous spatial point processes using case-control data. {\it Biometrics}, {\bf 63}, 550-557.
\bibitem{fan1993}
Fan, J. (1993). Local linear regression smoothers and their minimax efficiency. {\it Annals of Statistics}, {\bf 21}, 196-216.
\bibitem{friedman1981}
Friedman, J.H. and Stuetzle, W. (1981). Projection pursuit regression. {\it Journal of the American Statistical Association}, {\bf 76}, 817-823.
\bibitem{gu2013}
Gu, C. (2013). {\it Smoothing Spline ANOVA Models, 2nd Edition}. Springer, New York. 
\bibitem{guan2008}
Guan, Y. (2008). A KPSS test for stationarity for spatial point processes. {\it Biometrics}, {\bf 64}, 800-806.
\bibitem{guan2009}
Guan, Y. (2009). On nonparametric variance estimation for second-order statistics of inhomogeneous spatial point processes with known parametric intensity form. {\it Journal of the American Statistical Association}, {\bf 104}, 1482-1491.
\bibitem{guan2010}
Guan, Y. and Shen, Y. (2010). A weighted estimating equation approach for inhomogeneous spatial point processes. {\it Biometrika}, {\bf 97}, 867-880.
\bibitem{golub1979}
Golub, G.H., Heath, M., and Wahba, G. (1979). Generalized cross-validation as a method for choosing a good ridge parameter. {\it Technometrics}, {\bf 21}, 215-223.
\bibitem{hastie1990}
Hastie, T.J. and Tibshirani, R.J. (1990). {\it Generalized Additive Models}. Chapman and Hall, Washington, DC.
\bibitem{henrys2009}
Henrys, P.A. and Brown, P.E. (2009). Inference for cluster inhomogeneous spatial point processes. {\it Biometrics}, {\bf 65}, 423-430.
\bibitem{herrndorf1984}
Herrndorf, N. (1984). A functional central limit theorem for weakly dependent sequence of random variables. {\it Annals of Probability}, {\bf 12}, 141-153.
\bibitem{ibragimov1962}
Ibragimov, I.A. (1962). Some limit theorems for stationary processes. {\it Stochastic Processes and Their Applications}, {\bf 12}, 171-186.
\bibitem{ivanoff1982}
Ivanoff, G. (1982). Central limit theorems for point processes. {\it Stochastic Processes and Their Applications}, {\bf 12}, 171-186.
\bibitem{kimeldorf1971}
Kimeldorf, G. and Wahba, G. (1971). Some results on tchebycheffian spline functions. {\it Journal of Mathematical Analysis and Applications}, {\bf 33}, 82-94.
\bibitem{moller2007}
M{\o}ller, J. and Waagepetersen, R.P. (2007). Modern statistics for spatial point processes (with discussion). {\it Scandinavian Journal of Statistics}, {\bf 34}, 685-711.
\bibitem{ogata1988}
Ogata, Y. (1988). Statistical models for earthquake occurrences and residual analysis for point processes. {\it Journal of the American Statistical Association}, {\bf 83}, 9-27.
\bibitem{peng2005}
Peng, R.D., Schoenberg, F.P, and Woods, J.A. (2005). A space-time conditional intensity model for evaluating a wildfire hazard index. {\it Journal of the American Statistical Association}, {\bf 100}, 26-35.
\bibitem{pitman2007}
Pitman, A.J., Narisma, G.T., and McAneney, J. (2007). The impact of climate change on the risk of forest and grassland fires in Australia. {\it Climatic Change}, {\bf 84}, 383-401.
\bibitem{podur2002}
Podur, J., Martell, D.L., and Knight, K. (2002). Statistical quality control analysis of forest fire activity in Canada. {\it Canadian Journal of Forest Research}, {\bf 32}, 195-205.
\bibitem{ripley1976}
Ripley, B.D. (1976). The second-order analysis of spatial point processes. {\it Journal of Applied Probability}, {\bf 23}, 255-266.
\bibitem{rosenblatt1956}
Rosenblatt, M. (1956). A central limit theorem and a strong mixing condition. {\it Proceedings of the National Academy of Sciences of the United States of America}, {\bf 42}, 43-47. 
\bibitem{schoenberg2004}
Schoenberg, F.P. (2004). Testing separability in spatial-temporal marked point processes. {\it Biometrics}, {\bf 60}, 471-481.
\bibitem{stoyan1994}
Stoyan, D. and Stoyan, H. (1994). {\it Fractals, Random Shapes and Point Fields}. New York: Wiley.
\bibitem{stoyan1996}
Stoyan, D. and Stoyan, H. (1996). Estimating pair correlation functions of planar cluster processes. {\it Biometrical Journal}, {\bf 38}, 259-271.
\bibitem{vandervaart1998}
van der Vaart, A.W. (1998). {\it Asymptotic Statistics}, Cambridge University Press, Cambridge, UK.
\bibitem{waagepetersen2007}
Waagepetersen, R. (2007). An estimating function approach to inference for inhomogeneous Neyman-Scott process. {\it Biometrics}, {\bf 63}, 252-258. 
\bibitem{waagepetersen2009}
Waagepetersen,, R. and Guan, Y. (2009). Two-step estimation for inhomogeneous spatial point processes. {\it Journal of the Royal Statistical Society B}, {\bf 71}, 685-702.
\bibitem{wahba1990}
Wahba, G. (1990). Spline models for observational data. {\it CBMS-NSF Regional Conference Series in Applied Mathematics}, SIAM.
\bibitem{whitt2007}
Whitt, W. (2007). Proofs of the martingale FCLT. {\it Probability Surveys}, {\bf 4}, 268-302.
\bibitem{xiao2007}
Xiao, J. and Zhuang, Q. (2007). Drought effects on large fire activities in Canadian and Alaskan forests. {\it Environmental Research Letters}, {\bf 2}, 044003.
\bibitem{zhang2014a}
Zhang, T. and Zhuang, Q. (2014). On the local odds ratio between points and marks in marked point processes. {\it Spatial Statistics}, {\bf 9}, 20-37.
\bibitem{zhang2014c}
Zhang, T. and Zhou, B. (2014). Test for stationarity for spatial point processes in an arbitrary region. {\it Journal of Agricultural, Biological and Environmental Statistics}, {\bf 19}, 387-404.
\bibitem{zhang2014d}
Zhang, T. (2014). A Kolmogorov-Smirnov type test for independence between marks and points of marked point processes. {\it Electronic Journal of Statistics}, {\bf 8}, 2557-2584. 
\end{thebibliography}
\end{document}